\documentclass[authoryear,a4paper,review]{elsarticle}
\usepackage{amsmath,amssymb,natbib,hyperref,setspace,rotating}

\journal{Finance Research Letters}
\usepackage[left=1in,right=1in,top=1.4in,bottom=1.2in]{geometry}

\hypersetup{
	pdftitle={The financial value of the within-government political network: Evidence from Chinese municipal corporate bonds},
	pdfauthor={Jaehyuk Choi, Lei Lu, Heungju Park, Sungbin Sohn},
	pdfkeywords={},
	colorlinks=true,
	linkcolor=red,
	citecolor=blue,
	urlcolor=blue,
	bookmarksnumbered=true,
	pdfstartview=
}

\date{January 4, 2022}

\begin{document}

\begin{frontmatter}

\title{The financial value of the within-government political network:\\Evidence from Chinese municipal corporate bonds}

\author[phbsadd]{Jaehyuk Choi}
\ead{jaehyuk@phbs.pku.edu.cn}
\author[phbsadd]{Lei Lu}
\ead{lulei@sz.pku.edu.cn}
\author[skkadd]{Heungju Park\corref{corrauthor}}
\ead{hj.park@skku.edu}
\author[sogangadd]{Sungbin Sohn}
\ead{sungbsohn@sogang.ac.kr}
\cortext[corrauthor]{Corresponding author}
\address[phbsadd]{Peking University HSBC Business School, Nanshan, Shenzhen 518055, China}
\address[skkadd]{Sungkyunkwan University SKK Business School, Jongno-gu, Seoul 03063, Republic of Korea}
\address[sogangadd]{Sogang University School of Economics, Mapo-gu, Seoul 04107, Republic of Korea}

\begin{abstract}
This paper examines the effect of the political network of Chinese municipal leaders on the pricing of municipal corporate bonds. Using municipal leaders' working experience to measure the political network, we find that this network reduces the bond issuance yield spreads by improving the credit ratings of the issuer, the local government financing vehicle. The relationship between political networks and issuance yield spreads is strengthened in areas where financial markets and legal systems are less developed.
\end{abstract}


\end{frontmatter}

\section*{Highlights}
\begin{itemize}
\item We study a within-government political network using municipal corporate bonds.
\item Political network is measured by municipal leaders' upper-level-government guanxi.
\item Political network improves LGFV's credit rating, reducing their borrowing costs.
\item Political network is more valuable where market and legal systems are less developed.
\end{itemize}

\section{Introduction} \label{sec:intro} \noindent 
The impact of political connections on financial markets is a much-studied topic in academia~\citep{agrawal2001outside,khwaja2005lenders, claessens2008political,boubakri2008political,boubakri2012political,butler2009corruption,duchin2012politics,kusnadi2019political,klusak2020personal}. In particular, research using Chinese data has received much attention because China is a relation-centered society, and guanxi (personal networks in Chinese) plays a crucial role in lubricating business and helps companies compete for resources in both the private and public sectors~\citep{park2001guanxi,li2008political,mcnally2011china,tsai2019political,li2020what,li2020political,gao2021subnational}.

Unlike previous studies that have mostly focused on the connection between the government and corporations, this paper examines political networks within the Chinese government. Specifically, we investigate whether Chinese municipal leaders' political networks, measured by their working experience in upper-level (provincial or central) governments, affect the issuance pricing of municipal corporate bonds (MCBs). The seemingly contradictory term ``MCB'' comes from the fact that although these bonds are issued by a nongovernment entity called the local government financing vehicle (LGFV) in a strict legal sense, they are implicitly backed by municipalities. For this reason, we hypothesize that MCBs are highly susceptible to municipal leaders' political networks when securing financial resources to avoid default.

Using hand-collected data on municipal leaders' working experience, we find that the political network reduces the credit risks and issuance yield spreads of LGFVs. To test endogeneity and robustness, we also use a subsample of municipalities that have gained or lost political connections, a propensity score-matched sample, and municipality-year-level panel data. The results remain consistent in three alternative samples. Further analyses show that in regions where financial markets and legal systems are less developed, the political network has a stronger effect on the issuance yield spreads of MCBs. A policy change that attempts to remove the implicit guarantee has little impact on the negative association between the network and yield spread.

This paper contributes to the literature on MCBs and Chinese shadow banking~\citep{liu2017implicit,chen2020financing} by identifying political connections as a pricing determinant.\footnote{Notably, several studies employ political connections. For example, \citet{gao2021subnational} document the selective default of LGFVs' bank loans rather than MCBs. \citet{qian2018citc} investigates the effect of the anti-corruption campaign on LGFVs and MCBs but reports an insignificant impact on the yield spread.} Given that the risk profile of off-balance sheet financial products is inherently more obscure, noneconomic factors, such as political power, presumably play additional roles in the issuance pricing of shadow banking products. Our findings support this presumption.

\section{Institutional Background} \label{sec:inst} \noindent
The LGFV and MCBs have interesting policy backgrounds and are the keys to understanding concerns regarding the mounting debt of China's local governments~\citep{chen2020financing}. The birth of the LGFV has primarily been attributed to the tax-sharing system reform and budget law that was enacted in 1994, which reallocated a large portion of tax revenue from local governments to the central government and prohibited local governments from issuing debt. In response, local governments started using the LGFV to finance large infrastructure development off their balance sheets. The use of the LGFV exploded in 2009 as a vital channel for implementing the four trillion yuan economic stimulus package to cope with the economic slowdown after the 2008 global financial crisis. As the main target of the stimulus was local infrastructure development, the central government encouraged municipal governments to use LGFVs, circumventing the 1994 budget law. An MCB is a bond issued by an LGFV and, along with bank loans, is a primary source of LGFV funding. In particular, local governments started heavily issuing MCBs in 2012 to refinance mature bank loans~\citep{chen2020financing}.

Given the above background and China's long history of a centrally planned economy, it is not surprising that investors widely believe that the government will bailout LGFVs in the case of default and consider MCBs safe. However, which government (i.e., municipal, provincial, or central) is liable for the MCBs issued by the municipal LGFVs and to what extent it is liable are unclear. In fear of the growing risk of LGFVs and the government's guarantee being taken for granted, the State Council issued No.~43 Document in 2014 to prohibit local governments from raising debt via LGFVs and providing a guarantee for existing LGFV debt. Although this regulation's effect was virtually negated by the subsequent orders due to debt rollover pressure, it is an important regulation change to test the persistence of the political network's impact.

Therefore, how government bailout expectations are reflected in MCB pricing is an ongoing research question. \citet{ang2018great}, assuming that the central government will eventually back MCBs, focus on macroeconomic variables. \citet{liu2017implicit} argue that the fiscal conditions of the municipal (and to some extent, provincial) government became important pricing determinants after the first LGFV default in 2011. In line with this research, we investigate municipal leaders' political networks.

\section{Data} \label{sec:data} \noindent
We collect the yield data of 1,552 MCBs issued by 217 municipalities between 2009 and 2017 from the WIND database. Although MCBs are also issued by provinces, we focus on municipal MCBs, as we are concerned with upward political connections. We collect the yield of China Development Bank (CDB) bonds or Chinese government bonds with the same maturity to measure the MCB issuance yield spread. We construct municipal leaders' political network by manually collecting information from the People's Republic of China Official Records, People's Daily, and Xinhuanet. Specifically, municipal leaders form a political network ($Network = 1$) if the municipal Communist Party Committee Secretary has experience working for (1) the central or provincial government, (2) the central or provincial People's Congress, or (3) the central or provincial People's Political Consultative Conference. The People's Congress is China's law-making body, and the People's Political Consultative Conference has official advisory powers. Otherwise, the leader has no political network ($Network = 0$). Our sample includes 329 municipal leaders from related municipalities.

We consider variables for macroeconomic conditions (one-year Chinese government bond yield, PMI, M2, CPI, and stock market index) and local economic conditions (municipal-level GDP per capita, municipal government financial soundness, marketization index, and legalization index) to control for the issuing price of MCBs. We also control for issuer characteristics (size, leverage, ROE, growth rate of operating revenues, and solvency rate), leader characteristics (age and education) and bond characteristics (issuance amount, maturity, external guarantee, and credit ratings). See Table~\ref{tb:var} for the list of control variables and their sources.

Table~\ref{tb:summary} reports the descriptive statistics of key variables for the full sample and a propensity score-matched sample. To mitigate potential endogeneity concerns, we use the propensity score matching method. Specifically, we predict the probability of a municipality being politically networked using local economic conditions and issuer and leader characteristics. Then, we match each issuance observation in a municipality with a political network with that in a municipality without a political network, without replacement. The matched sample includes 638 bonds issued under a political network and 638 bonds issued under no network.

Panel A of Table~\ref{tb:summary} provides comparisons of various characteristics for subsamples by network and shows that the characteristics of municipalities with political networks differ substantially from those of municipalities without networks in various dimensions, which suggests that unobservable omitted variables may influence both political network and other issue characteristics. After propensity score matching, the descriptive statistics are presented in Panel B of Table~\ref{tb:summary}. Most of the significant mean differences across politically networked and non-networked issues become small and insignificant, alleviating selection bias concerns.

\section{Empirical Results} \label{sec:res} \noindent
Our baseline regression is as follows:
\begin{equation} \label{eq:1}
YS = \beta_0 + \beta_{1} Network +  \gamma\, Control +\varepsilon ,
\end{equation}
where $YS$ refers to the issuance yield spread of an MCB. $Network$ is the indicator of the political network of the leader in the municipality that issued the MCB. $Control$ includes the control variables for economic conditions, the characteristics of municipalities, issuers (LGFVs), leaders, and bonds (MCBs), and the issuance year of MCBs. All right-hand-side variables are lagged.

The results are reported in Table~\ref{tb:main}. First, using a full sample in Panel A, we find that municipal leaders' political network significantly reduces the issuance yield spreads. On average, municipalities with political networks issue MCBs with yields 18 basis points lower than those without networks. The results are robust to the two yield spread definitions, $YS_{CDB}$ and $YS_{GovBond}$. The previous studies on political connections have documented that political leaders have exerted discretionary powers to selectively default on debt and the firms with political connections have preferential access to resources \citep{claessens2008political, guo2014political, gao2021subnational}. Our finding shows that the same underlying mechanism applies to the within-government political network as well as to the government-corporate connection. The estimates of the other control variables are generally consistent with previous research~\citep{liu2017implicit}. Notably, the yield spread is negatively correlated with the fiscal health of the LGFV ($Asset$) and municipality ($Local Financing Gap$).

One may argue that a leader's political connection is endogenous to the municipality characteristics that influence the MCB yield spread. In China, municipal leaders are appointed by the Communist Party, and politically networked leaders may be favorably appointed to municipalities whose fiscal condition is sound or improving. In fact, approximately half of issue observations are from municipalities where the network status remains unchanged through leadership changes. Considering that one municipality has approximately seven party secretaries, on average, during our sample period, this suggests that the appointment is not purely random.

To address this selection issue, we use the subsample of municipalities where there has been at least one transition of the leader's political network. Panel B shows that even after dropping the municipalities with persistent network status, our results remain unaffected. We further alleviate the endogeneity concern by re-estimating our regression models with a propensity score-matched sample described in Table~\ref{tb:summary}. Since politically networked municipalities are matched to nonnetworked municipalities that are otherwise identical, the endogeneity problem associated with municipal characteristics can be reduced. Panel~C shows that the significance and magnitude of the $Network$ coefficient barely change. Together with the fact that the independent variable is lagged by a month, the findings suggest that the lower MCB issuance yield spread in politically networked municipalities is driven neither by observable differences in municipalities nor by unobservable selection bias.

To confirm the robustness of our findings, we restructure the issue-level cross-sectional data into municipality-year-level panel data where multiple bonds issued in the same municipality-year are represented by the average weighted by their issuance amounts.\footnote{The results with equally weighted averages are nearly identical to those with the amount-weighted averages.}  Then, we conduct a panel regression controlling for the year- and municipality-fixed effects. Panel D shows that our main finding remains unchanged.\footnote{The municipality-year panel setting also enables us to test the frequency or amount of the issuance as dependent variables. In unreported tables, however, we do not find a statistically significant effect of the political network, which rejects the demand-and-supply channel for reduced yield spreads. We thank an anonymous referee for the suggestion.}

Next, to identify the channel of reduced issuance yield spreads, we employ a structural model for mediation analysis~\citep{heckman2013understanding,heckman2015econometric}. Using LGFVs' credit ratings as a mediator variable, we can infer the direct and indirect effects of the political network. Table~\ref{tb:mediation} shows the following simultaneous equation analysis performed on the subsamples with network status changes (Panel A) and propensity score-matched samples (Panel B):
\begin{equation}\label{eq:2}
\begin{aligned}
Credit       &= \delta_0 + \delta_{1} Network +  \theta Control + \eta \\
YS &= \beta_0 + \beta_{1} Network  + \beta_{2} Credit + \gamma\, Control  + \varepsilon,
\end{aligned}
\end{equation}
where $Credit$ is the credit rating of the LGFV that issues the MCB. Controls include the same variables used in Equation~\eqref{eq:1}. The results are robust to the two sample sets. Specifically, Columns (1) and (3) of each panel show that the political network significantly enhances the credit ratings of LGFVs. In Columns (2) and (4) of each panel, the issuance yield spreads decrease with both the political network and credit ratings. Our results support the hypotheses that the political network enhances the credit ratings of LGFVs and that these enhanced credit ratings, in turn, serve as an effective channel for reducing the borrowing costs of municipalities. The standalone effect of the political network is still significant, implying that the political network can explain the MCB yield spreads beyond the effect of the credit improvement channel.

Furthermore, we conduct several additional analyses with the propensity score-matched sample.\footnote{In untabulated tests, we also examine full and subsamples with changes in the network status and find that our results are robust to the choices of the samples.} First, when the political network is decomposed into networks with the central and provincial governments (Column (1) in Table~\ref{tb:robust}), we find that the network's impact on the issuance price of MCBs is mainly driven by the municipality--province connection. This finding is not only intuitive but also consistent with \citet{liu2017implicit}. Since the central government tries to reduce outstanding municipal MCBs by switching to more transparent provincial municipal bonds, \citet{liu2017implicit} argue that the fiscal conditions of provinces have become important determinants of MCB pricing. As the swap program is managed with strict provincial quotas, we naturally expect that the municipality--province connection helps secure the quota.

Second, we examine whether regional heterogeneities in institutionalization influence the political network's strength. Guanxi plays a larger role in the absence of institutional support ~\citep{xin1996guanxi}, and circumstances such as the lack of liquidity and disclosure are considered a major impediment of the MCB market~\citep{lam2019china}. Therefore, we expect that the influence of the political network is more pronounced in regions with underdeveloped financial markets and legal systems. Specifically, we test the interaction between the political network and marketization and legalization indices. Columns (2) and (3) in Table~\ref{tb:robust} show that the interaction terms have significantly positive coefficients, consistent with our hypothesis. The negative standalone impacts of the indices on the issuance yield spreads are also consistent with \citet{lam2019china}, as the issuance is costly in underdeveloped market environments.

Finally, we examine whether there are significant differences in the political network's effect before and after No.~43 Document as an important regulation change. The Chinese government announced this regulation in September 2014 to prohibit LGFVs from being used as financing platforms and to end the relationship between LGFVs and local governments. However, the regulation allows LGFVs to issue MCBs for repaying existing bank loans or other borrowings. In fact, the MCB issuance increased, not decreased, after No.~43 Document, implying a limited impact on the MCB market. Our empirical findings are generally consistent with this reality. Despite its apparent aim, this regulation has little effect on the issue spread. Specifically, the insignificant interaction term in Column (4) of each panel implies that the regulation has had little effect on the relationship between the political network and the issuance price of MCBs. Chinese investors seem to continue to value political networks despite government intentions.

\section{Conclusion} \label{sec:conc} \noindent
This study examines the impact of municipal leaders' political networks on the issuance pricing of MCBs. Specifically, we analyze whether the political network can give more assurance to the MCBs issued by the associated LGFVs and find that the political network reduces the credit risks of these LGFVs, leading to a reduction in their issuance yield spreads. The network that matters for the borrowing cost of MCBs is the connection to the provincial government rather than to the central government. Furthermore, in regions where the market and legal systems are less developed, the political network has a stronger effect on the issuance yield spreads of MCBs. We also find evidence that No.~43 Document, which was introduced to curb the ``incorrect'' investor perception of implicit guarantees on MCBs, has had little effect in that its introduction has not significantly changed the influence of the political network. Last, despite our best efforts, we acknowledge that the potential endogeneity concern between political networks and the issuance pricing of MCBs may still remain. Employing an exogenous shock to lead the unexpected changes in municipal leaders' political network can overcome our limitation regarding the endogeneity issue and be fruitful for future work.

\singlespacing
\bibliography{ChinaMCB}


\clearpage
\newpage
\begin{sidewaystable}[htbp]
\caption{\textbf{Descriptive Statistics for Full Sample and Propensity Score-Matched Sample}}\label{tb:summary}
\smallskip\footnotesize This table presents descriptive statistics of key variables for the full sample and propensity score-matched sample. The variables are defined in Table~\ref{tb:var}. The t-statistics of the differences in all variables between non-networked and networked groups are shown. ***, **, and * denote significance at the 1\%, 5\%, and 10\% levels, respectively.
\begin{center}
\begin{tabular}{lccccccccccccc}
\hline \hline

                        &\multicolumn{6}{c}{Panel A: Full Sample} &&\multicolumn{6}{c}{Panel B: Propensity Score-Matched Sample}\\
                        \cline{2-7} \cline{9-14}
	                    &\multicolumn{2}{c}{Without network}&&\multicolumn{2}{c}{With network}&&&\multicolumn{2}{c}{Without network}&&\multicolumn{2}{c}{With network}&\\
                        \cline{2-3} \cline{5-6} \cline{9-10} \cline{12-13}
Variables	            &Mean	&Median		&&Mean	&Median	&Mean Diff	&&Mean	&Median		   &&Mean	&Median	&Mean Diff \\
\hline
Bond Variables          &      	&		    &&	    &	    &		    &&	    &		       &&		&       &          \\
$Maturity$	            &1.95	&1.95		&&1.95	&1.95	&0.00		&&1.95	&1.95		   &&1.95	&1.95	&0.01      \\
$Issue Amounts$ 	    &11.56	&11.51		&&11.60	&11.61	&-0.04*		&&11.56	&11.51		   &&11.59	&11.61	&-0.03     \\
$Guarantee$ 	        &0.26	&0.00		&&0.24	&0.00	&0.02		&&0.27	&0.00		   &&0.24	&0.00	&0.03      \\
Issuer Variables        &       & 		    &&	    &	    &		    &&	    &		       &&		&	    &          \\
$Current Ratio$         &6.30	&4.48		&&6.94	&4.58	&-0.63*		&&6.75	&4.65		   &&6.94	&4.58	&-0.19     \\
$EBITDA$                &0.20	&0.10		&&0.17	&0.08	&0.03*		&&0.17	&0.10		   &&0.17	&0.08	&0.00      \\
$Leverage$ 	            &39.91	&40.08		&&40.86	&41.33	&-0.95		&&41.03	&40.84		   &&40.88	&41.33	&0.15      \\
$ROE$	                &3.68	&3.29		&&3.29	&2.82	&0.39***	&&3.46	&3.14		   &&3.31	&2.83	&0.15      \\
$Growth$	            &29.47	&6.29		&&27.30	&9.04	&2.17		&&28.79	&4.56		   &&27.61	&9.37	&1.18      \\
$Asset$	                &14.05	&14.02		&&14.22	&14.17	&-0.16***	&&14.18	&14.14		   &&14.21	&14.15	&-0.03     \\
City Variables          &      	&	    	&&	    &	    &		    &&	    &		       &&		&       &          \\
$Municipal GDP$         &10.62	&10.57		&&10.72	&10.72	&-0.10***	&&10.68	&10.67		   &&10.71	&10.72	&-0.03     \\
$Local Financing Gap$   &0.54	&0.51		&&0.55	&0.51	&-0.01		&&0.54	&0.52		   &&0.55	&0.52	&-0.01     \\
Leader Variables    	&	    &		    &&	    &	    &		    &&	    &		       &&	    &       &          \\
$Age$	                &3.98	&3.97		&&3.96	&3.97	&0.01***	&&3.97	&3.97		   &&3.97	&3.97	&0.00      \\
$Education$	            &1.96	&2.00		&&2.12	&2.00	&-0.16***	&&2.05	&2.00		   &&2.10	&2.00	&-0.06     \\
Observations	        &\multicolumn{2}{c}{904}&&\multicolumn{2}{c}{648}&&&\multicolumn{2}{c}{638}&&\multicolumn{2}{c}{638}&\\
\hline\hline
\end{tabular}
\end{center}
\end{sidewaystable}


\clearpage
\newpage
\begin{sidewaystable}[htbp]
\caption{\textbf{The Impact of Political Network on Issuance Yield Spreads}}\label{tb:main}
\smallskip\footnotesize This table presents regression results of issuance yield spreads on political network. The results with full sample are reported in Panel A. In Panel B, we report the regression results for the subsample of municipalities with changes in the network status, while the results with propensity score-matched sample are reported in Panel C. Panel D reports estimates of municipality-year-level panel regressions of issuance yield spreads on political network with city and year fixed effects and bond issuance amounts are used as the weights for value weighted averages in Panel D. The dependent variable is issuance yield spreads and all independent variables are defined in Table~\ref{tb:var}. The t-statistics reported in parentheses are based on heteroscedasticity-robust standard errors clustered by municipality. ***, **, and * denote significance at the 1\%, 5\%, and 10\% levels, respectively.
\begin{center}
\begin{tabular}{lccccccccccc}
\hline \hline
               &\multicolumn{2}{c}{Panel A: Full sample}& &\multicolumn{2}{c}{Panel B: Changes in network}& &\multicolumn{2}{c}{Panel C: PS-matched}& &\multicolumn{2}{c}{Panel D: Municipality-year level sample}\\
\cline{2-3} \cline{5-6} \cline{8-9} \cline{11-12}
                     &  $YS_{CDB}$&$YS_{GovBond}$& &  $YS_{CDB}$&$YS_{GovBond}$&  &  $YS_{CDB}$&$YS_{GovBond}$&  &  $YS_{CDB}$&$YS_{GovBond}$\\
\hline
$Network$	         &   -0.187***&  -0.181***&  &   -0.187***&   -0.183***&  &   -0.182***&   -0.171***&  &  -0.299***	&  -0.311***    \\
                     &     (-4.06)&    (-3.84)&  &     (-2.87)&     (-2.70)&  &     (-3.81)&     (-3.53)&  &  (-3.85)	&  (-4.18)    \\
$Maturity$ 	         &     0.580**&    0.548**&  &       0.268&       0.304&  &     0.660**&     0.636**&  &   0.088	&  0.098    \\
                     &      (2.53)&     (2.46)&  &      (0.57)&      (0.67)&  &      (2.49)&      (2.48)&  &  (0.29)	&  (0.32)    \\
$Issue Amounts$      &   -0.191***&  -0.225***&  &     -0.156*&    -0.195**&  &   -0.175***&   -0.212***&  &  -0.266***	&  -0.286***    \\
                     &     (-3.39)&    (-3.77)&  &     (-1.85)&     (-2.16)&  &     (-3.05)&     (-3.38)&  &  (-2.71)	&  (-2.83)    \\
$Guarantee$	         &      -0.102&     -0.080&  &       0.030&       0.057&  &     -0.122*&      -0.093&  &  -0.155*	&  -0.149*    \\
                     &     (-1.61)&    (-1.30)&  &      (0.33)&      (0.64)&  &     (-1.78)&     (-1.39)&  &  (-1.86)	&  (-1.76)    \\
$Current Ratio$      &      -0.001&     -0.002&  &       0.002&      -0.000&  &      -0.004&      -0.005&  &  -0.008	&  -0.009    \\
                     &     (-0.45)&    (-0.71)&  &      (0.42)&     (-0.04)&  &     (-1.44)&     (-1.43)&  &  (-1.15)	&  (-1.36)    \\
$EBITDA$	         &       0.056&      0.045&  &       0.016&       0.023&  &      0.131*&       0.113&  &   0.094	&  0.035    \\
                     &      (0.86)&     (0.64)&  &      (0.22)&      (0.26)&  &      (1.66)&      (1.31)&  &  (0.74)	&  (0.29)    \\
$Leverage$	         &      -0.001&     -0.001&  &       0.000&       0.000&  &      -0.001&      -0.001&  &   0.001	&  0.002    \\
                     &     (-0.47)&    (-0.57)&  &      (0.01)&      (0.17)&  &     (-0.53)&     (-0.64)&  &  (0.34)	&  (0.48)    \\
$ROE$		         &      -0.011&     -0.007&  &      -0.015&      -0.012&  &   -0.032***&    -0.024**&  &  -0.011	&  -0.008    \\
                     &     (-1.11)&    (-0.70)&  &     (-1.03)&     (-0.81)&  &     (-2.83)&     (-2.01)&  &  (-0.67)	&  (-0.45)    \\
$Growth$	         &      -0.000&     -0.000&  &       0.000&       0.000&  &      -0.000&      -0.000&  &  -0.000	&  -0.000    \\
                     &     (-0.43)&    (-0.56)&  &      (0.93)&      (1.09)&  &     (-1.04)&     (-1.03)&  &  (-0.37)	&  (-0.07)    \\
$Asset$		         &   -0.147***&  -0.125***&  &   -0.201***&   -0.184***&  &   -0.167***&   -0.140***&  &  -0.158**	&  -0.158**    \\
                     &     (-3.35)&    (-2.73)&  &     (-3.08)&     (-2.72)&  &     (-3.78)&     (-2.92)&  &  (-2.55)	&  (-2.60)    \\
$Municipal GDP$      &       0.043&      0.033&  &      -0.009&      -0.027&  &       0.058&       0.048&  &  -0.279**	&  -0.303**    \\
                     &      (0.74)&     (0.57)&  &     (-0.12)&     (-0.35)&  &      (0.93)&      (0.77)&  &  (-2.47)	&  (-2.47)    \\
$Local Financing Gap$&   -0.531***&  -0.512***&  &      -0.334&      -0.324&  &   -0.515***&   -0.489***&  &  -0.525	&  -0.602    \\
                     &     (-3.55)&    (-3.36)&  &     (-1.44)&     (-1.33)&  &     (-3.18)&     (-2.96)&  &  (-1.20)	&  (-1.40)    \\
$Age$	             &      -0.290&     -0.366&  &     -0.924*&     -0.997*&  &      -0.205&      -0.334&  &  -0.551	&  -0.776    \\
                     &     (-0.82)&    (-1.01)&  &     (-1.70)&     (-1.77)&  &     (-0.50)&     (-0.79)&  &  (-0.80)	&  (-1.15)    \\
$Education$          &      -0.004&     -0.018&  &      0.066*&       0.054&  &       0.020&       0.009&  &  -0.029	&  -0.015    \\
                     &     (-0.11)&    (-0.53)&  &      (1.70)&      (1.33)&  &      (0.51)&      (0.24)&  &  (-0.49)	&  (-0.25)    \\
$PMI$	             &     0.076**&     0.048*&  &       0.073&       0.041&  &       0.033&       0.015&  &  0.120**	&  0.105**    \\
                     &      (2.49)&     (1.72)&  &      (1.41)&      (0.85)&  &      (0.89)&      (0.41)&  &  (2.39)	&  (2.23)    \\
$M2$	             &    -0.064**&   -0.059**&  &   -0.094***&   -0.094***&  &    -0.051**&     -0.043*&  &  -0.135***	&  -0.117***    \\
                     &     (-2.58)&    (-2.43)&  &     (-2.60)&     (-2.71)&  &     (-1.99)&     (-1.69)&  &  (-3.21)	&  (-2.73)    \\
$CPI$	             &    0.304***&   0.391***&  &     0.378**&    0.480***&  &    1.028***&    0.965***&  &   0.336	&  0.354*    \\
                     &      (2.83)&     (3.64)&  &      (2.37)&      (3.06)&  &      (4.89)&      (4.94)&  &  (1.62)	&  (1.66)    \\
$Stock$	             &      -0.000&     -0.000&  &      -0.000&      -0.000&  &       0.000&      -0.000&  &  -0.001	&  -0.001    \\
                     &     (-0.28)&    (-0.36)&  &     (-0.81)&     (-0.71)&  &      (0.06)&     (-0.14)&  &  (-0.72)	&  (-0.83)    \\
$1Yr Treasury$       &   -0.287***&  -0.127***&  &   -0.255***&      -0.095&  &   -0.188***&      -0.034&  &  -0.362***	&  -0.191**    \\
                     &     (-5.86)&    (-2.61)&  &     (-4.04)&     (-1.57)&  &     (-3.20)&     (-0.58)&  &  (-4.25)	&  (-2.21)    \\
Constant             &  -27.247***& -34.090***&  &   -30.722**&   -38.686**&  &  -99.914***&  -91.735***&  &  -25.307	&  -25.039    \\
                     &     (-2.65)&    (-3.30)&  &     (-2.02)&     (-2.57)&  &     (-4.84)&     (-4.80)&  &  (-1.22)	&  (-1.18)    \\
Fixed Effects        &        Year&       Year&  &        Year&        Year&  &         Year&       Year&  &Municipality \& Year&Municipality \& Year\\
Adjusted $R^{2}$        &       0.532&      0.620&  &       0.516&       0.607&  &       0.496&       0.591&  &  0.627	    &  0.694     \\
Observations         &       1,552&      1,552&  &         733&         733&  &       1,276&       1,276&  &  621	    &  621       \\
\hline\hline
\end{tabular}
\end{center}
\end{sidewaystable}


\clearpage
\newpage
\begin{sidewaystable}[htbp]
\caption{\textbf{Mediation Analysis with Credit Risk}}\label{tb:mediation}
\smallskip\footnotesize This table presents the impacts of credit risks on the relationship between the political network and issuance yield spreads with subsample with changes in the network status and propensity score-matched sample. We present the results of mediation analysis with credit rating scores. The dependent variable is issuance yield spreads and all independent variables are defined in Table~\ref{tb:var}. The t-statistics reported in parentheses are based on heteroscedasticity-robust standard errors clustered by municipality. ***, **, and * denote significance at the 1\%, 5\%, and 10\% levels, respectively.
\begin{center}
\begin{tabular}{lccccccccccc}
\hline \hline
&  \multicolumn{5}{c}{Panel A: Subsample with changes in network status} & &\multicolumn{5}{c}{Panel B: PS-matched sample} \\
\cline{2-6} \cline{8-12}
                        &  $Credit $& $YS_{CDB}$& &$Credit $& $YS_{GovBond}$& & $Credit $& $YS_{CDB}$& &$Credit $ & $YS_{GovBond}$ \\
\cline{2-3} \cline{5-6} \cline{8-9} \cline{11-12}
                        &        (1)&     (2)   & &      (3)&         (4)	& &       (1)&        (2)& &      (3)&         (4)\\
\hline
$Network$	&   0.12***&    -0.13**& &  0.12***&      -0.12* & &   0.18***&    -0.09**& &  0.18***&      -0.08*\\
                        	&     (2.62)&    (-2.13)& &   (2.63)&     (-1.93)  & &  (5.17)&    (-2.10)& &   (5.20)&     (-1.82)\\
$Credit$		&    &   -0.42***& &         &    -0.44***  & &               &   -0.53***& &         &    -0.54***\\
                        	&          &    (-7.64)& &         &     (-7.86)  & &         &   (-11.61)& &         &    (-11.27)\\
$Control_{Bond}$&         No&        Yes& &       No&         Yes& &        No&        Yes& &       No&         Yes\\
$Control_{Issuer}$      &        Yes&        Yes& &      Yes&         Yes& &       Yes&        Yes& &      Yes&         Yes\\
$Control_{City}$        &        Yes&        Yes& &      Yes&         Yes& &       Yes&        Yes& &      Yes&         Yes\\
$Control_{Leader}$      &         No&        Yes& &       No&         Yes& &        No&        Yes& &       No&         Yes\\
$Control_{Macro}$       &         No&        Yes& &       No&         Yes& &        No&        Yes& &       No&         Yes\\
$Control_{Year}$        &        Yes&        Yes& &      Yes&         Yes& &       Yes&        Yes& &      Yes&         Yes\\
Observations            &    733&        733& &      733&         733  & &       1,276&      1,276& &    1,276&       1,276\\
\hline\hline
\end{tabular}
\end{center}
\end{sidewaystable}


\clearpage
\newpage
\begin{table}[htbp]
\caption{\textbf{Political Network and Issuance Yield Spreads: Robustness}}\label{tb:robust}
\smallskip\footnotesize This table presents the relationship between political connections and issuance yield spreads with propensity score-matched sample. The dependent variable is issuance yield spreads and all independent variables are defined in Table~\ref{tb:var}. The t-statistics reported in parentheses are based on heteroscedasticity-robust standard errors clustered by municipality. ***, **, and * denote significance at the 1\%, 5\%, and 10\% levels, respectively.
\begin{center}
\begin{tabular}{lccccccccc}
\hline \hline
                           &\multicolumn{4}{c}{Panel A: $YS_{CDB}$}& &\multicolumn{4}{c}{Panel B: $YS_{GovBond}$}\\
\cline{2-5} \cline{7-10}
                           &         (1)&         (2)&         (3)&         (4)& &        (1)&         (2)&         (3)&         (4)\\
\hline
$Network_{Province}$       &    -0.19***&            &            &            & &   -0.18***&            &            &            \\
                           &     (-3.90)&            &            &            & &    (-3.66)&            &            &            \\
$Network_{Central}$        &       -0.04&            &            &            & &      -0.01&            &            &            \\
                           &     (-0.27)&            &            &            & &    (-0.04)&            &            &            \\
$Network$                  &            &    -0.19***&    -0.19***&    -0.19***& &           &    -0.18***&    -0.18***&    -0.18***\\
                           &            &     (-4.26)&     (-4.07)&     (-3.41)& &           &     (-3.87)&     (-3.71)&     (-2.97)\\
$Network$ $\times$ $Market$&            &      0.06**&            &            & &           &      0.06**&            &            \\
                           &            &      (2.60)&            &            & &           &      (2.21)&            &            \\
$Market$                   &            &    -0.10***&            &            & &           &    -0.10***&            &            \\
                           &            &     (-5.05)&            &            & &           &     (-4.95)&            &            \\
$Network$ $\times$ $Legal$ &            &            &     0.02***&            & &           &            &      0.02**&            \\
                           &            &            &      (3.16)&            & &           &            &      (2.52)&            \\
$Legal$                    &            &            &    -0.02***&            & &           &            &    -0.02***&            \\
                           &            &            &     (-3.71)&            & &           &            &     (-3.07)&            \\
$Network$ $\times$ No.~43   &            &            &            &        0.02& &           &            &            &        0.01\\
                           &            &            &            &      (0.27)& &           &            &            &      (0.14)\\
No.~43                      &            &            &            &       -0.23& &           &            &            &    -0.70***\\
                           &            &            &            &     (-1.42)& &           &            &            &     (-4.29)\\
$Control_{Bond}$           &         Yes&         Yes&         Yes&         Yes& &        Yes&         Yes&         Yes&         Yes\\
$Control_{Issuer}$         &         Yes&         Yes&         Yes&         Yes& &        Yes&         Yes&         Yes&         Yes\\
$Control_{City}$           &         Yes&         Yes&         Yes&         Yes& &        Yes&         Yes&         Yes&         Yes\\
$Control_{Leader}$         &         Yes&         Yes&         Yes&         Yes& &        Yes&         Yes&         Yes&         Yes\\
$Control_{Macro}$          &         Yes&         Yes&         Yes&         Yes& &        Yes&         Yes&         Yes&         Yes\\
$Control_{Year}$           &         Yes&         Yes&         Yes&         Yes& &        Yes&         Yes&         Yes&         Yes\\
Adjusted $R^{2}$              &       0.497&       0.510&       0.503&       0.498& &      0.592&       0.601&       0.595&       0.608\\
Observations               &       1,276&       1,276&       1,276&       1,276& &      1,276&       1,276&       1,276&       1,276\\
\hline\hline
\end{tabular}
\end{center}
\end{table}


\clearpage
\newpage
\begin{sidewaystable}[htbp]
\renewcommand\thetable{A1}
\caption{\textbf{Variable Definitions}}\label{tb:var}
\smallskip \scriptsize
\begin{center}
\begin{tabular}{lll}
\hline \hline
Variable Name	           &  Definition		                                                                                                 &   Sources             \\
\hline
$YS_{CDB}$                 &  Yield difference between MCB and CDB bond with the same maturity           		                                 &   WIND                \\
$YS_{GovBond}$               &  Yield difference between MCB and Chinese government bond with same maturity	 	                                 &   WIND                \\
$Network$	               &  Dummy variable equal to 1 if the municipal secretary has work experience in upper-level government and 0 otherwise &   Manual Collection    \\
$Maturity$	               &  Log of the debt maturity		                                                                                     &   WIND                 \\
$Issue Amounts$	           &  Log of the issuance amount		                                                                                 &   WIND                 \\
$Guarantee$	               &  Dummy variable equals to 0 if there is no external guarantee by third parties, and 1 otherwise                       &   WIND                 \\
$Current Ratio$	           &  Current ratio of the issuing company		                                                                         &   CSMAR                \\
$EBITDA$	               &  EBITDA divided by interest-bearing debt		                                                                     &   CSMAR                \\
$Leverage$	               &  Total liability divided by total asset		                                                                     &   CSMAR                \\
$ROE$	                   &  Return on equity		                                                                                             &   CSMAR                \\
$Growth$	               &  Annual growth rate of operating revenues		                                                                     &   CSMAR                \\
$Asset$	                   &  Log of the issuing company's asset		                                                                         &   WIND                 \\
$Municipal GDP$	           &  Log of GDP per capita in the municipality		                                                                     &   Local Finance Bureau \\
$Local Financing Gap$      &  Government revenue divided by government expenditure of the fiscal year prior to the issuance date                 &   Local Finance Bureau \\
$Age$                      &  Log of municipal secretary's age                                                                                   &   CSMAR                \\
$Education$                &  Municipal secretary's education level of below undergraduate, undergraduate, postgraduate/MBA and Ph.D.              &   CSMAR                \\
$PMI$	                   &  Purchasing managers' index		                                                                                 &   NBSPRC               \\
$M2$	                   &  Monthly year-on-year growth of broad money 		                                                                 &   NBSPRC               \\
$CPI$	                   &  Consumer price index 		                                                                                         &   NBSPRC               \\
$Stock$	                   &  Daily change of Shanghai Composite Index 		                                                                     &   WIND                 \\
$1Yr Treasury$	           &  Yield of one-year Chinese government bond		                                                                     &   WIND                 \\
$Credit$	               &  Firm credit rate when the bond is firstly issued; A value of 1 assigned to A- and 6 assigned AAA		             &   WIND                 \\
$Market$	               &  Marketization index		                                                                                         &   \citet{fan2016neri} \\
$Legal$	                   &  Legalization index		                                                                                         &   \citet{fan2016neri} \\
\hline\hline
\end{tabular}
\end{center}
\end{sidewaystable}

\end{document}